\documentclass{article}

\usepackage{arxiv}

\usepackage[utf8]{inputenc} 
\usepackage[T1]{fontenc}    
\usepackage{hyperref}       
\usepackage{url}            
\usepackage{booktabs}       
\usepackage{amsfonts}       
\usepackage{nicefrac}       
\usepackage{microtype}      
\usepackage{lipsum}		
\usepackage{setspace}
\usepackage{graphicx}
\usepackage{doi}
\usepackage{amsmath}
\usepackage[numbers,sort&compress]{natbib}

\title{Goldene monolayer as a highly effective catalyst for polysulfide anchoring and conversion: A theoretical study}

\author{%
\parbox{0.95\linewidth}{\centering
\href{https://orcid.org/0000-0001-7653-0428}{\includegraphics[scale=0.09]{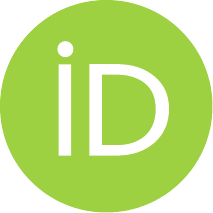}\hspace{1mm}}Nicolas F.~Martins\textsuperscript{1},
\href{https://orcid.org/0000-0002-8366-7227}{\includegraphics[scale=0.09]{icons/orcid.pdf}\hspace{1mm}}Jos\'e A.~dos S.~Laranjeira\textsuperscript{1},
\href{https://orcid.org/0000-0001-5048-0696}{\includegraphics[scale=0.09]{icons/orcid.pdf}\hspace{1mm}}Bill D.~A.~Huacarpuma\textsuperscript{2}
\href{https://orcid.org/0000-0003-4699-5886}{\includegraphics[scale=0.09]{icons/orcid.pdf}\hspace{1mm}}Kleuton A.~L.~Lima\textsuperscript{3},
\href{https://orcid.org/0000-0001-7468-2946}{\includegraphics[scale=0.09]{icons/orcid.pdf}\hspace{1mm}}Luiz A.~Ribeiro Jr\textsuperscript{3,$\dag$},
and
\href{https://orcid.org/0000-0002-5217-7145}{\includegraphics[scale=0.09]{icons/orcid.pdf}\hspace{1mm}}Julio R. Sambrano\textsuperscript{1},
 \\
\vspace{0.6em}
{\normalfont\normalsize
\textsuperscript{1}Modeling and Molecular Simulation Group, S\~ao Paulo State University (UNESP), School of Sciences, Bauru 17033-360, SP, Brazil\\
\textsuperscript{2}Computational Materials Laboratory, LCCMat, Institute of Physics, University of Bras\'ilia, 70910-900, Bras\'ilia, Federal District, Brazil\\
\textsuperscript{3}Department of Applied Physics and Center for Computational Engineering and Sciences, State University of Campinas, Campinas, 13083-859, SP, Brazil\\
\vspace{0.6em}
\href{https://scholar.google.com/citations?user=EBMMpbYAAAAJ&hl=pt-BR}{\includegraphics[scale=0.05]{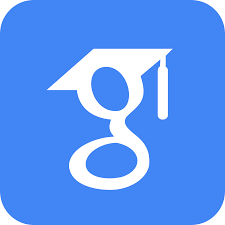}} \href{https://www.linkedin.com/in/nicolas-martins-765136388/}{\includegraphics[scale=0.05]{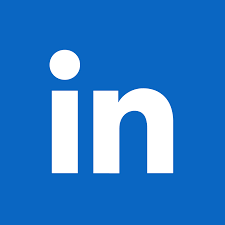}}\hspace{0.1cm}\texttt{\textsuperscript{*}nicolas.ferreira@unesp.br } \\
\vspace{0.1cm}
\href{https://scholar.google.com/citations?user=Gs2UkTgAAAAJ&hl=pt-BR}{\includegraphics[scale=0.05]{icons/gscholar.png}} \href{https://www.linkedin.com/in/julio-sambrano-0257591a5/}{\includegraphics[scale=0.05]{icons/linkedin.png}}\hspace{0.1cm}\texttt{\textsuperscript{$\dag$}jr.sambrano@unesp.br }\\
}
}%
}

\renewcommand{\shorttitle}{Nicolas F.~Martins et al.}

\hypersetup{
pdftitle={3d-dodeca},
pdfsubject={3d-dodeca},
pdfauthor={Kleuton A. L. Lima, Luiz A. Ribeiro Jr},
pdfkeywords={Two-dimensional metals, Kagome lattice, Goldene, Lattice dynamics, Strain engineering, First-principles calculations.},
}

\begin{document}
\maketitle

\onehalfspacing

\begin{abstract}
We use first-principles density functional theory to investigate how lithium sulfide and polysulfide clusters (Li$_2$S, Li$_2$S$_2$, Li$_2$S$_4$, Li$_2$S$_6$, Li$_2$S$_8$, and S$_8$) bind to Goldene, a new two-dimensional gold allotrope. All Li–S species exhibit robust binding to Goldene. The adsorption energies range from $-4.29$ to $-1.90$~eV. S$_8$ that is alone interacts much less strongly. Charge density difference and Bader analyses indicate that substantial charge is transferred to the substrate, with a maximum $0.92\,e$ for Li-rich clusters. This transfer induces polarization at the interface and shifts the work function to 5.30-5.52 eV. Projected density-of-states calculations indicate that Au-$d$ and S-$p$ states strongly mix near the Fermi level. This hybridization indicates that the electronic coupling is strong. Based on these results, the reaction free-energy profile for the stepwise conversion of S$_8$ to Li$_2$S on Goldene is thermodynamically favorable. The overall stabilization is -3.64 eV, and the rate-determining barrier for the Li$_2$S$_2$ $\rightarrow$ Li$_2$S step is 0.47 eV. This shows that Goldene is an effective surface for anchoring and mediating lithium polysulfide reactions.
\end{abstract}


\section{Introduction}

Because of their low environmental impact, high theoretical energy density, and the natural abundance of sulfur, lithium-sulfur (Li-S) electrochemical systems have attracted significant interest \cite{guo2024interface, liu2021electrolyte}. The dissolution and migration of lithium polysulfides (Li$_2$S$_x$, $2 \leq x \leq 8$) and interfacial instability during cycling are two intrinsic challenges that limit practical implementation despite these benefits \cite{huang2022recent}. Low Coulombic efficiency and rapid capacity fading are caused by these issues, especially the uncontrolled movement of soluble polysulfides between electrodes \cite{wang2023recent, chen2022insight}. As a result, interfacial engineering becomes an essential tactic for improving performance. 

Two-dimensional (2D) materials are increasingly used as platforms for anchoring polysulfide and mediating reactions because they offer large surface areas, can be tuned to alter their electronic structure, and can react with a wide range of chemicals \cite{jana2020rational, fan2022two, muhammad2024first, martins2025anchoring}. Graphene-based materials were among the first studied in this context because they exhibit high electrical conductivity and mechanical stability \cite{yang2021graphene, al2022two, martins2025exploring}. However, because pristine graphene is nonpolar, lithium polysulfides do not adhere well to it \cite{hou2016design}. To address this limitation, several functionalization strategies have been proposed to enhance chemical affinity, but they often come at the cost of increased structural complexity or reduced conductivity \cite{Kuila2012,Chen2012}. 

On the other hand, transition-metal dichalcogenides such as MoS$_2$ have polar surfaces that can form stronger chemical bonds with lithium polysulfides \cite{eng2019controlled, yang2024environmental}, since exposed metal sites increase the adsorption energies and catalytic activity in polysulfide conversion reactions. The MXenes family is another class of nanomaterials explored as potential anchoring surfaces because they are good electrical conductors, have hydrophilic surfaces, and exhibit a strong affinity for sulfur species \cite{zhao2021status, zhang2021mxene}. Liang and coworkers \cite{liang2022fluorine} found that applying a fluorine-free fabrication of Ti$_3$C$_2$ MXenes reduces the environmental impact of MXene synthesis and also promotes electron reservoirs due to the reactive oxygen species on the terminated surface. From a catalytic perspective, Fang and colleagues \cite{fang2023effective} executed a screening of 27 potential MXene cathodes for Li-S batteries based on density functional theory (DFT). They obtained Ti$_2$CS$_2$, Mo$_2$CS$_2$, and W$_2$CS$_2$ as promising catalysts due to the less endothermic Gibbs free energies during the rate-determining step from Li$_2$S$_2$ to Li$_2$S. Other studies have also examined the effect of S- or Se-terminated MXenes on polysulfide conversion \cite{martins2025investigating, upadhyay2025unraveling}.

The recent synthesis of Goldene, a 2D monolayer allotrope of gold, has created new possibilities for designing interfacial materials \cite{kashiwaya2024synthesis}. As a metallic 2D material with high electrical conductivity, structural robustness, and exposed Au atoms, Goldene offers new possibilities for a myriad of applications. Specifically, Kumar and coworkers \cite{kumar2025co2} proposed the functionalization of the Goldene monolayer with different metal adatoms to enhance the Co$_2$ activation. Sheremetyeva and Meunier, for example, found that Goldene supports a high coverage ratio of adsorbed hydrogen atoms, which could amplify research in the field of catalysis and hydrogen storage with Au-based surfaces. Different theoretical efforts have also been executed to investigate an overview of the main structural and electronic properties of Goldene \cite{pereira2025does, dos2025exploring, mortazavi2024Goldene}. However, there are no reports on the potential use of the Goldene monolayer as an anchoring material to enhance polysulfide conversion. Additionally, it is well documented that Au-based nanomaterials strongly interact with sulfur-containing species \cite{zou2024tailoring, sun2024metal, chao2025leveraging}. This indicates that Goldene can enhance the charge–discharge mechanism of Li–S batteries by combining the promising properties of the noble metal Au with the unique attributes of 2D materials.
 
In this work, we present a systematic first-principles investigation of the adsorption, electronic structure, and reaction energetics of lithium sulfide and polysulfide clusters (Li$_2$S$_n$, where n = 1, 2, 4, 6, and 8) on Goldene. By combining adsorption energy analysis, charge density difference (CDD), work-function modulation, projected density of states, and reaction free-energy profiling, we elucidate the fundamental mechanisms governing Li–S interactions at the Goldene interface. Our results position Goldene as a promising metallic 2D platform for controlling polysulfide chemistry and reaction pathways at the nanoscale.

\section{Computational Methods}

All calculations were conducted using density functional theory (DFT) as executed in the Vienna ab initio Simulation Package (VASP) \cite{Kresse1999}. The projector-augmented-wave (PAW) method \cite{Blchl1994} was used to describe the interactions between valence electrons and ionic cores. Simultaneously, exchange-correlation effects were accounted for using the generalized gradient approximation within the Perdew-Burke-Ernzerhof (PBE) framework \cite{perdew_1_1991, Perdew1996}. To ensure that the total energies and electronic properties were correct, a plane-wave kinetic energy cutoff of 520 eV was used. 

Goldene was modeled as a 2D slab with periodic boundary conditions applied in the in-plane directions. A vacuum spacing of 20~\r{A} was introduced along the out-of-plane direction to avoid spurious interactions between periodic images. Structural optimizations were carried out using a $\Gamma$-centered $6 \times 6 \times 1$ Monkhorst--Pack $k$-point mesh, while a denser $10 \times 10 \times 1$ grid was adopted for electronic structure and projected density of states (PDOS) calculations. Long-range dispersion interactions were accounted for using the DFT-D3 method \cite{grimme2010consistent}.

All atomic positions were fully relaxed until the total energy variation between ionic steps was below $10^{-5}$~eV and the residual Hellmann--Feynman forces on each atom were smaller than 0.01~eV/\r{A}. S$_8$ and Li$_2$S$_n$ (n = 1, 2, 4, 6, and 8) clusters were initially placed above the Goldene surface in multiple orientations, and the most stable adsorption geometries were identified through total energy minimization.

The adsorption energy ($E_{\mathrm{ads}}$) of each cluster on Goldene was calculated using the following equation:
\begin{equation}
E_{\mathrm{ads}} = E_{\mathrm{Goldene+cluster}} - E_{\mathrm{Goldene}} - E_{\mathrm{cluster}} .
\end{equation}

where $E_{\mathrm{Goldene+cluster}}$, $E_{\mathrm{Goldene}}$, and $E_{\mathrm{cluster}}$ represent the total energies of the combined system, the pristine Goldene monolayer, and the isolated Li--S cluster, respectively. Negative values of $E_{\mathrm{ads}}$ indicate energetically favorable adsorption.

The elementary reaction steps involved in the formation of a single Li$_2$S molecule are described as follows:
\begin{gather}
\mathrm{S_8^* + 2Li^+ + 2e^- \rightarrow Li_2S_8^*} \\
\mathrm{Li_2S_8^* \rightarrow Li_2S_6^* + \tfrac{1}{4}S_8} \\
\mathrm{Li_2S_6^* \rightarrow Li_2S_4^* + \tfrac{1}{4}S_8} \\
\mathrm{Li_2S_4^* \rightarrow Li_2S_2^* + \tfrac{1}{4}S_8} \\
\mathrm{Li_2S_2^* \rightarrow Li_2S^* + \tfrac{1}{8}S_8}
\end{gather}

Here, the superscript $^*$ denotes adsorption at an active site on the substrate. The Gibbs free energy change $\Delta G$ associated with each elementary step is calculated using the following expression:
\begin{equation}
\Delta G = \Delta E + \Delta E_{\mathrm{ZPE}} - T \Delta S ,
\label{eq:DG}
\end{equation}
where $\Delta E$ represents the adsorption energy, while $\Delta E_{\mathrm{ZPE}}$ and $T \Delta S$ correspond to the zero-point energy and entropy differences between products and reactants, respectively. These quantities were obtained from vibrational frequency calculations performed at 298~K. In this work, $\Delta E_{\mathrm{DFT}}$ was determined from the difference in total energies of the adsorbed intermediate states before and after each elementary reaction step. The terms $\Delta E_{\mathrm{ZPE}}$ and $\Delta S$ correspond to the variations in zero-point vibrational energy and entropy, respectively, evaluated at 298.15~K. Previous theoretical investigations on Li--S battery systems have shown that the contributions of $\Delta E_{\mathrm{ZPE}}$ and $T\Delta S$ are relatively small and can be safely neglected \cite{wu2025ds, yu2021exploring}. Consequently, the Gibbs free energy change can be reasonably approximated as $\Delta G \approx \Delta E_{\mathrm{DFT}}$.

\section{Results and Discussion}

\subsection{Structural description}

\begin{figure*}[!t]
    \centering
    \includegraphics[width=0.85\linewidth]{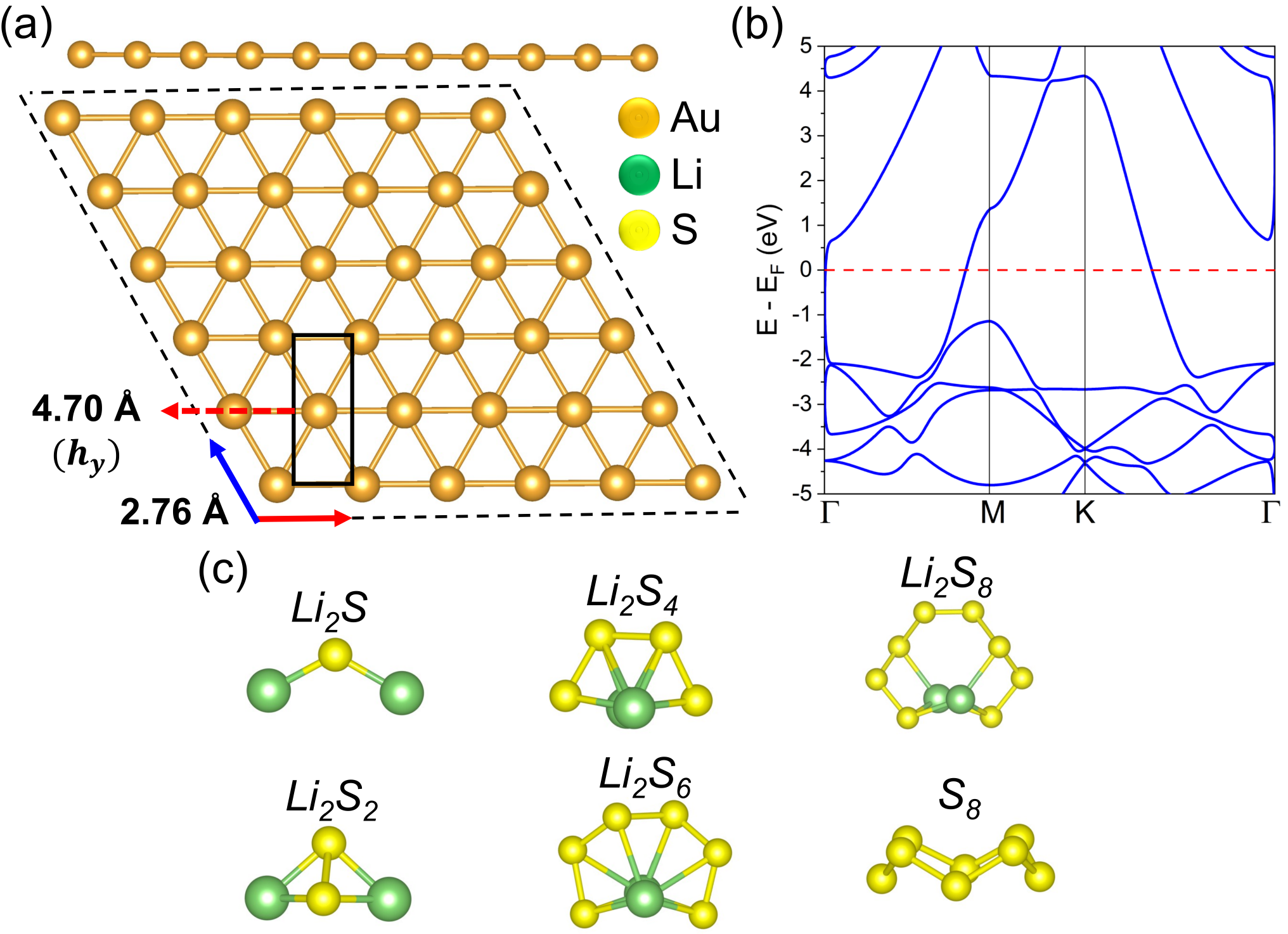}
    \caption{(a) Top and side views of the optimized Goldene monolayer and (b) its electronic band structure, with the Fermi level set to 0 eV. (c) Fully optimized isolated sulfur-containing species considered in this study, including Li$_2$S, Li$_2$S$_2$, Li$_2$S$_4$, Li$_2$S$_6$, Li$_2$S$_8$, and S$_8$.}
\label{fig:1}
\end{figure*}

Figure~\ref{fig:1} illustrates the structural models employed in this work. Panel (a) displays the optimized Goldene monolayer used as the two-dimensional substrate, with the in-plane lattice parameters highlighted as 2.76~\AA{} and 4.70~\AA{} along the indicated crystallographic directions. The dashed outline denotes the simulation cell adopted for the periodic calculations. Panel (b) presents the electronic band structure of the Goldene monolayer, confirming its intrinsic metallic character, as evidenced by multiple bands crossing the Fermi level (set to 0~eV). This metallicity is a key attribute for its investigation as a promising platform for Li–S batteries, since high electrical conductivity is essential to enhance charge–discharge kinetics and suppress the shuttle effect. Finally, panel (c) shows the isolated and fully optimized sulfur-containing species considered in this study—Li$_2$S, Li$_2$S$_2$, Li$_2$S$_4$, Li$_2$S$_6$, Li$_2$S$_8$, and S$_8$—which serve as molecular and cluster building blocks for the subsequent interfacial calculations.

\begin{figure*} 
    \centering
    \includegraphics[width=0.8\linewidth]{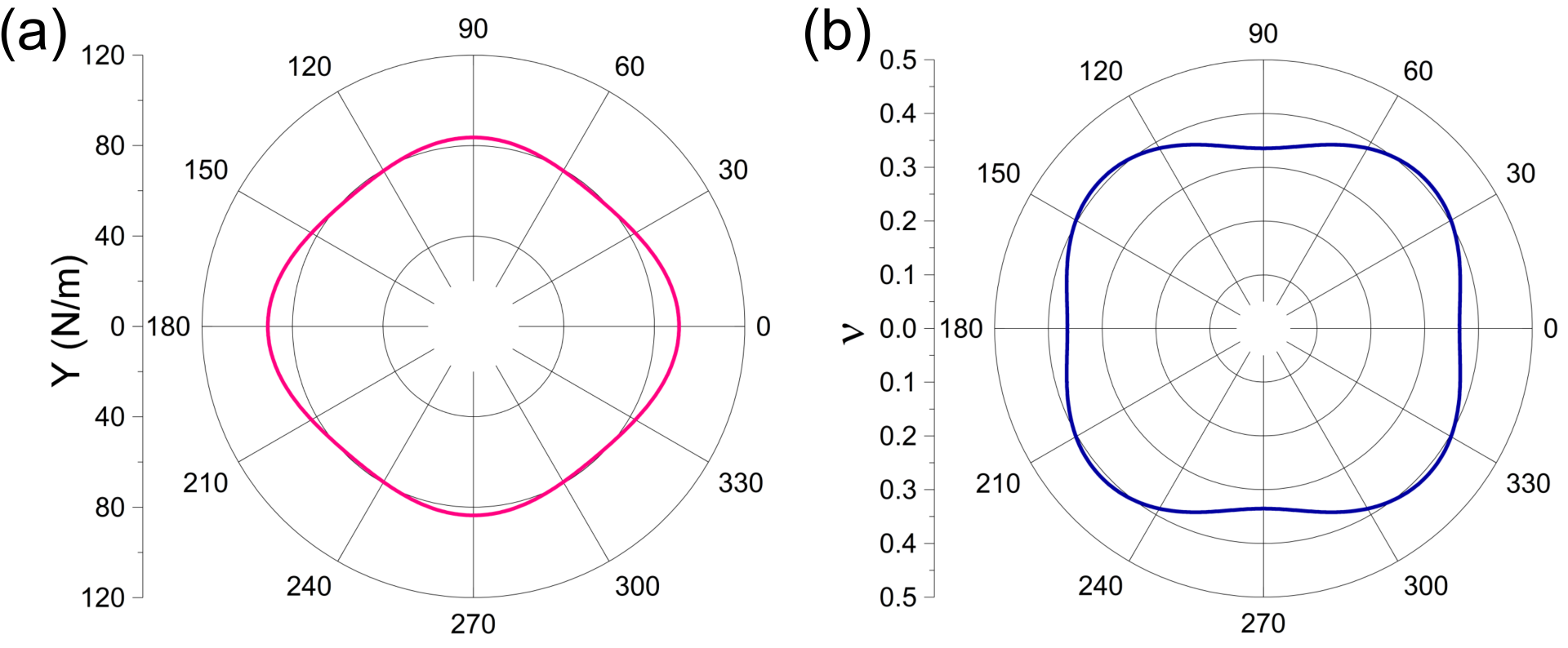}
    \caption{(a) Young's moduli and (b) Poisson's ratio for Goldene monolayer.}
    \label{fig:2}
\end{figure*}

We further evaluated the mechanical properties of the Goldene monolayer by calculating its elastic constants (C$_{11}$, C$_{12}$, C$_{22}$, and C$_{66}$), as well as the polar representations of Young’s modulus (Y) and Poisson’s ratio ($\nu$), as shown in Figure~\ref{fig:2}. The calculated stiffness constants are 103.56~N/m (C$_{11}$), 34.71~N/m (C$_{12}$), 95.22~N/m (C$_{22}$), and 28.11~N/m (C$_{66}$), indicating a higher mechanical rigidity compared to Goldene analogs such as copperene and silverene \cite{dos2025exploring}. The polar plots in Figure~\ref{fig:2} reveal a pronounced anisotropic mechanical behavior, with a maximum Young’s modulus of 90.90~N/m and a maximum Poisson’s ratio of 0.41. These results demonstrate that Goldene exhibits a robust mechanical response, highlighting its suitability for applications in energy storage systems.

\subsection{Interaction of S$_8$ and Li$_2$S$_n$ (n = 1, 2, 4, 6, and 8) clusters}

\begin{figure*}
    \centering
    \includegraphics[width=1\linewidth]{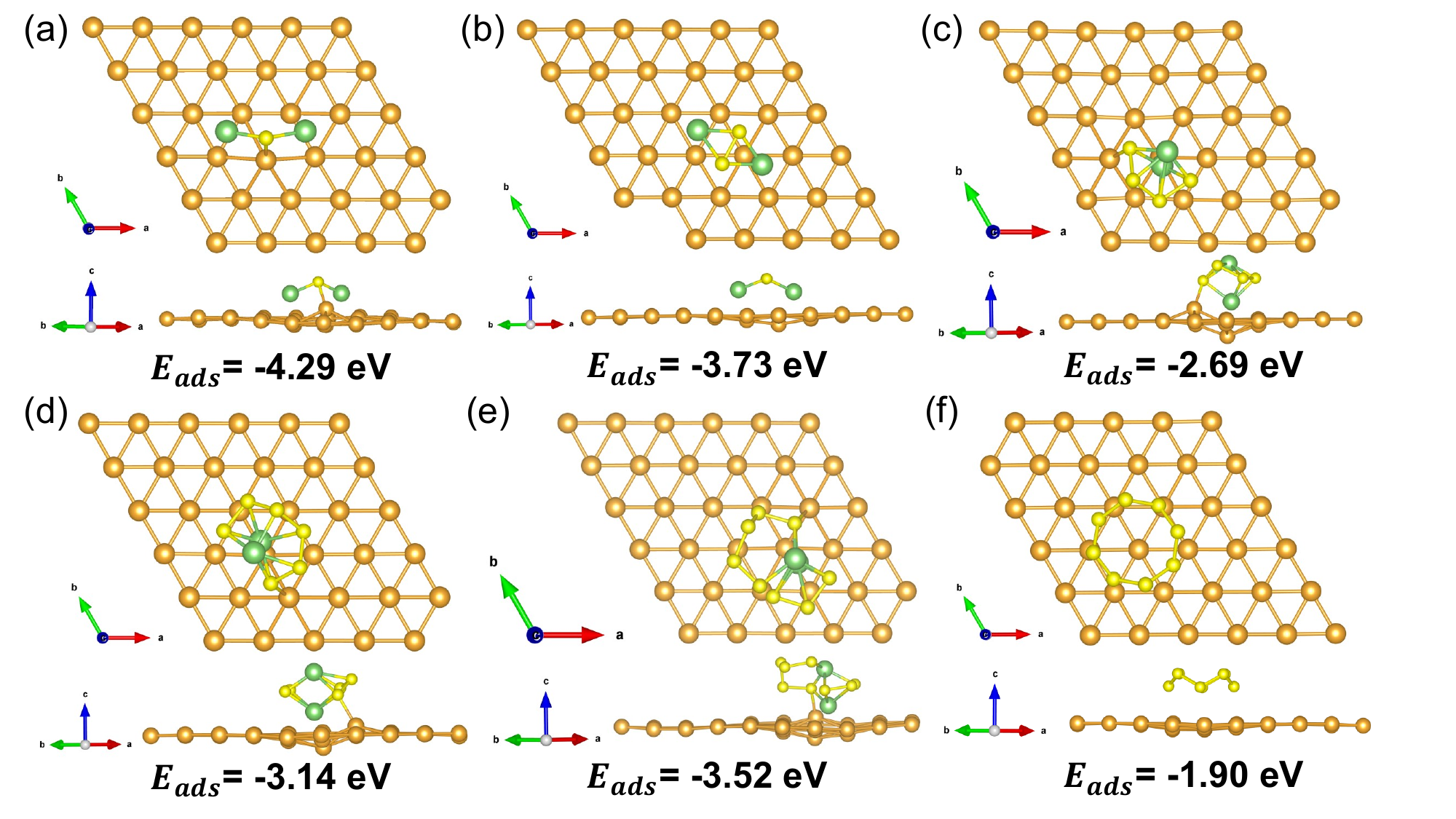}
    \caption{Optimized adsorption configurations of Li$_2$S$_n$ and S$_8$ clusters on Goldene, together with their corresponding adsorption energies ($E_{\mathrm{ads}}$). Panels (a)–(f) show the most stable geometries for Li$_2$S, Li$_2$S$_2$, Li$_2$S$_4$, Li$_2$S$_6$, Li$_2$S$_8$, and S$_8$, respectively, viewed from the top and side.}
    \label{fig:3}
\end{figure*}

To see if Goldene can work as a cathode material, we first look at how it interacts with sulfur-containing compounds. Figure~\ref{fig:3} shows the most stable ways in which lithium sulfide and lithium polysulfide clusters can adhere to the Goldene surface. All Li–S species create stable energy configurations, which means that Goldene can hold polysulfides in place over a wide range of chain lengths.  It is possible to identify that all Li$_2$S$_n$ clusters, including the S$_8$ molecule, maintain the molecular shape visualized in Figure \ref{fig:1}b. This finding is important since the cathode materials should capture the LiPS without dissociating or degrading the polysulfide species. 

Our results indicate that all LiPS are strongly adsorbed on the Goldene surface, suggesting the remarkable effect of the Au atom in interacting with the polysulfides to enhance the binding affinity. The adsorption energy results, also included in Figure \ref{fig:3}, corroborate the previous observation. The energy ranges from -1.90 to -4.29 eV, with the largest negative energy values being computed for Li$_2$S (E$_{ads}$ = -4.29 eV) and Li$_2$S$_2$ (E$_{ads}$ = -3.73 eV). Low-sulfur chains like Li$_2$S and Li$_2$S$_2$ are expected to present higher interaction energies due to the contribution of both Li atoms in the final anchoring (see Figures \ref{fig:3}a and b). On the other hand, the computed adsorption energies slightly decrease as the sulfur concentration increases. In general, the Goldene monolayer presents substantial binding strength for polysulfides compared to graphene and other graphene-based materials, which usually show adsorption energies lower than 2 eV (in absolute values) \cite{zeng2019single, velez2021role, li2025anchoring}.

\begin{figure*}
    \centering
    \includegraphics[width=0.75\linewidth]{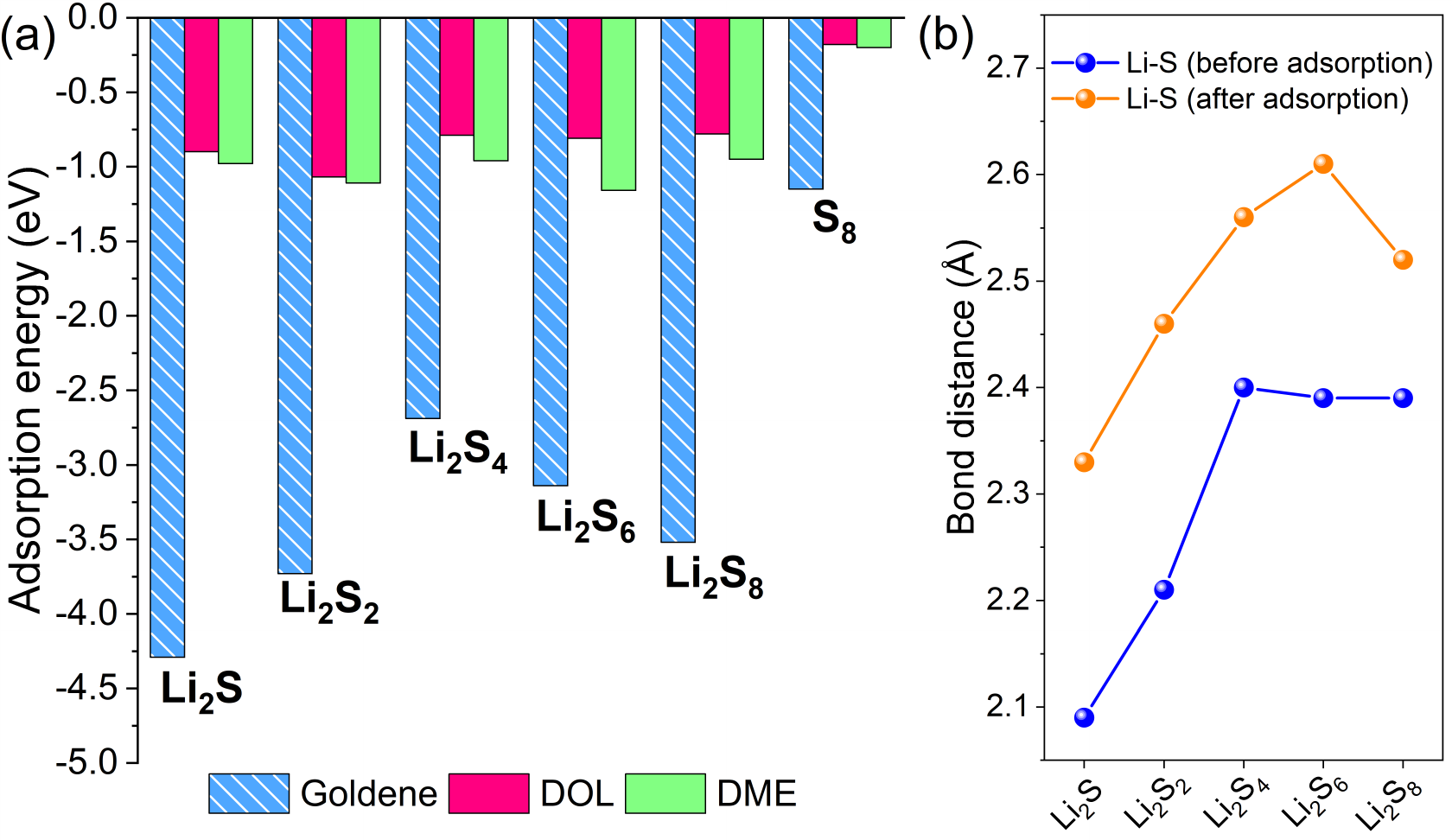}
    \caption{(a) Comparison of adsorption energies (E$_{ads}$) of Li$_2$S$_n$ and S$_8$ cluster species on Goldene and in common organic electrolyte solvents (DOL and DME), and the (b) Li-S bond distances before and after each lithium polysulfide adsorption on Goldene.}
    \label{fig:4}
\end{figure*}

From a practical point of view, it is important that lithium polysulfide binds strongly to the cathode surface so that it does not dissolve into the electrolyte. The adsorption energies of lithium polysulfides in common organic solvents such as 1,3-dioxolane (DOL) and 1,2-dimethoxyethane (DME) are typically between $-0.18$ and $-1.16$~eV \cite{yu2022bc6n}, which are much weaker than the values we found for Goldene (see Figure \ref{fig:4}). This comparison indicates that, upon lithiation, sulfur-containing species preferentially remain adsorbed on the Goldene surface rather than dissolve into the electrolyte, which is essential for reducing the shuttle effect and improving cycling stability in Li–S systems.

 This comparison indicates that, upon lithiation, sulfur-containing species preferentially remain adsorbed on the Goldene surface rather than dissolve into the electrolyte, which is essential for reducing the shuttle effect and improving cycling stability in Li–S systems. For rapid kinetics during polysulfide conversion, particularly in the formation of the final product (Li$_2$S$^{*}$), cleavage of the Li–S bond is required \cite{zhao2024kinetic, wang2025steric}. As shown in Fig.~\ref{fig:4}b, a clear stretching of the Li--S bond is observed for all Li$_2$S$_n$ clusters. This effect is more pronounced for low-sulfur species (Li$_2$S and Li$_2$S$_2$), where Li atoms migrate across the Goldene surface to delocalize electrons owing to their strong donor character. This behavior leads to elongation of the Li-S bond length and consequently promotes faster lithium kinetics.

\begin{figure*}
    \centering
    \includegraphics[width=0.75\linewidth]{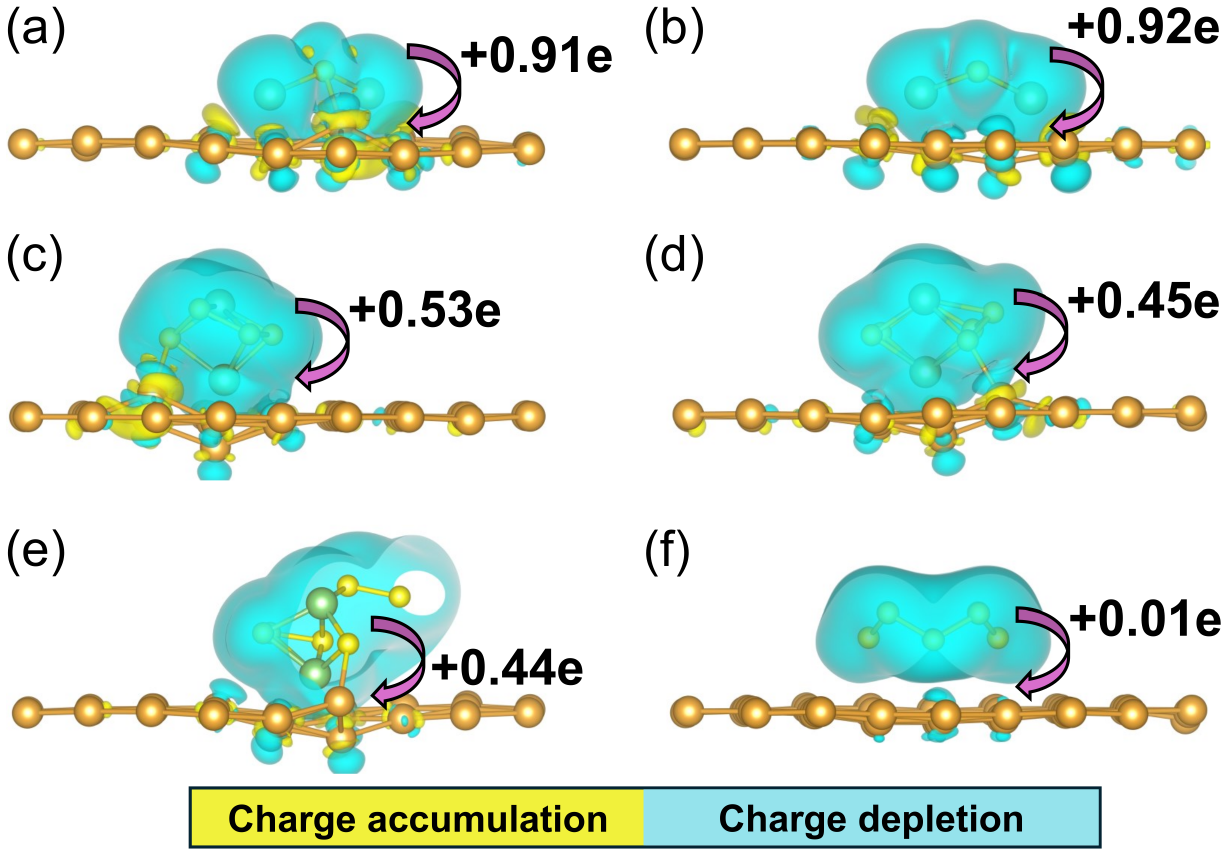}
    \caption{Charge density difference (CDD) plots for Li$_2$S$_n$ and S$_8$ clusters adsorbed on Goldene. Panels (a)–(f) correspond to Li$_2$S, Li$_2$S$_2$, Li$_2$S$_4$, Li$_2$S$_6$, Li$_2$S$_8$, and S$_8$, respectively. Yellow and cyan isosurfaces represent charge accumulation and depletion, respectively, induced by adsorption.}
    \label{fig:5}
\end{figure*}

To complement the structural results, Bader charge analysis and charge density difference (CDD) maps were computed for S$_8$ and all Li$_2$S$_n$ clusters. The CDD was obtained using the following expression:  \begin{equation}
\Delta \rho(\mathbf{r}) =
\rho_{\mathrm{Goldene+cluster}}(\mathbf{r})
- \rho_{\mathrm{Goldene}}(\mathbf{r})
- \rho_{\mathrm{cluster}}(\mathbf{r}) .
\end{equation}

Figure \ref{fig:5} shows the obtained CDD maps for each configuration along with the amount of charge transferred. This figure provides a direct view of how interfacial interactions alter the electronic structure. In all cases, there is a clear buildup of electron density at the interface between sulfur atoms and the Goldene surface. There are also clear charge depletion regions around lithium atoms in the clusters. This pattern shows that charge is moving from the Li–S species to the Goldene substrate. Bader charge analysis shows that the amount of charge transferred depends on the number of lithium atoms in the cluster and how they connect to the rest of the cluster. As expected, Li$_2$S and Li$_2$S$_2$ exhibit the greatest charge transfer, reaching about $0.91\,e$ and $0.92\,e$, respectively, indicating that these species are highly ionic and have strong electronic interactions with the Goldene surface. The transferred charge decreases as the sulfur chain lengthens: it ranges from $0.53\,e$ for Li$_2$S$_4$ to $0.45\,e$ for Li$_2$S$_6$ and $0.44\,e$ for Li$_2$S$_8$. This means that the charge is spreading out more within the polysulfide structure and that each lithium atom contributes less charge to the interface. On the other hand, the significant contribution of vdW forces on the S$_8$ adsorption induces an almost negligible charge transfer from this cluster to the Goldene monolayer. Finally, we can attest that our charge-transfer analysis is closely linked to the previous results concerning the binding strength of polysulfides on Goldene.

\begin{figure*}
    \centering
    \includegraphics[width=0.70\linewidth]{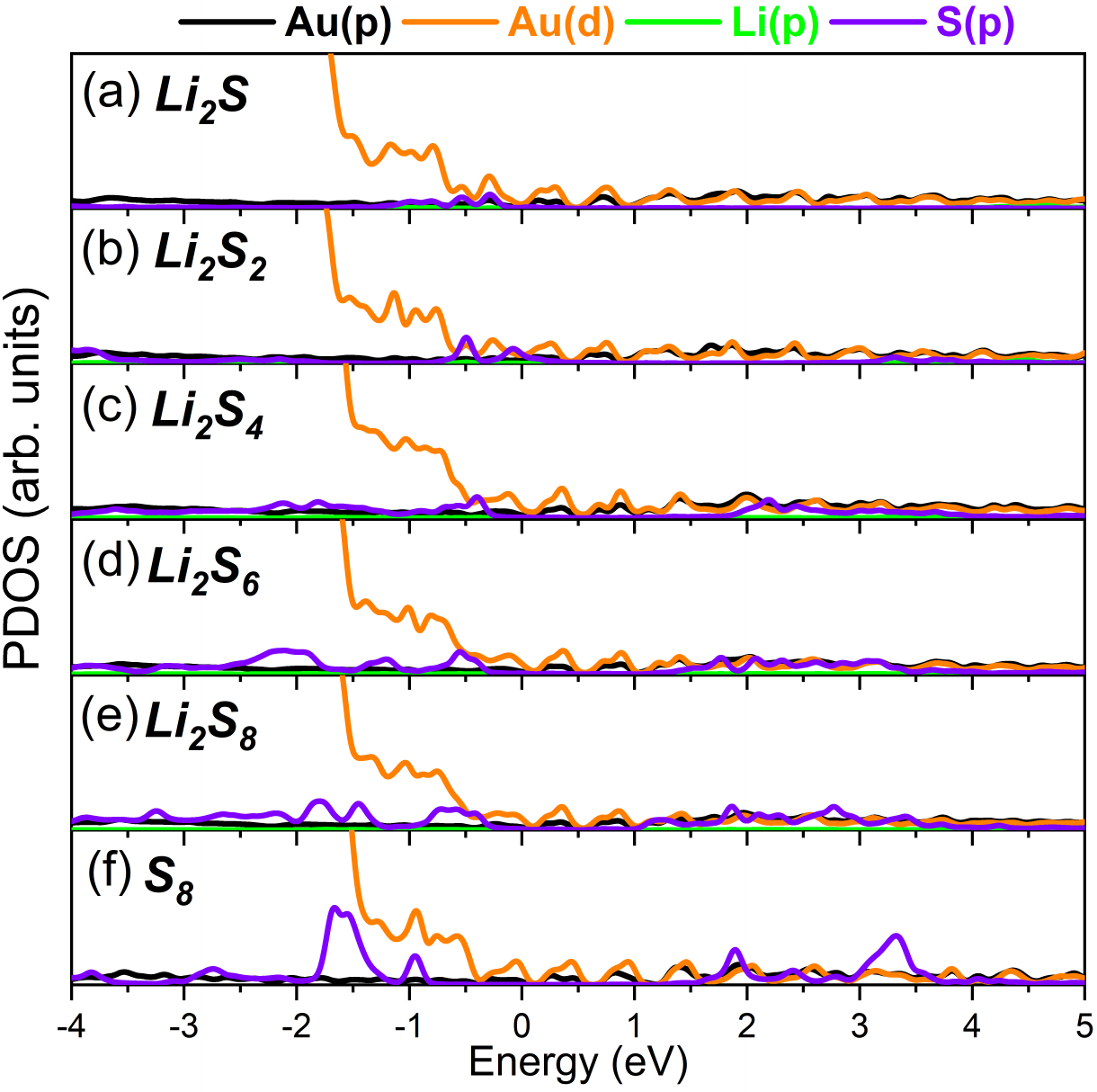}
    \caption{Projected density of states (PDOS) for Goldene with adsorbed lithium sulfide and polysulfide species. Panels (a)–(f) correspond to Li$_2$S, Li$_2$S$_2$, Li$_2$S$_4$, Li$_2$S$_6$, Li$_2$S$_8$, and S$_8$, respectively. Contributions from Au $p$ and $d$, Li $p$, and S $p$ orbitals are shown. The Fermi level is set to zero energy.}
    \label{fig:6}
\end{figure*}

To clarify the electronic interactions between sulfur-containing species and the Goldene substrate, projected density-of-states calculations were conducted, as illustrated in Figure\ref{fig:6}. The PDOS profiles show that Goldene maintains its metallic character in all cases, even though lithium sulfide and polysulfide clusters adhere to it well. There are still many Au $d$ states at the Fermi level, indicating that the adsorption process does not compromise the substrate's excellent electronic conductivity. This is an important requirement for cathode materials in Li–S systems. In Li-containing species, significant hybridization between Au $d$ and S $p$ states occurs near the Fermi level, especially in Li$_2$S and Li$_2$S$_2$. This overlap indicates that chemical Au–S bonds are forming, consistent with the short interfacial distances and high adsorption energies. As the length of the polysulfide chain increases, the contribution of S $p$ states near the Fermi level becomes more spread out. This is because sulfur orbitals in longer chains are less localized; yet, they still interact with the Goldene surface.

\begin{figure*}
    \centering
    \includegraphics[width=0.85\linewidth]{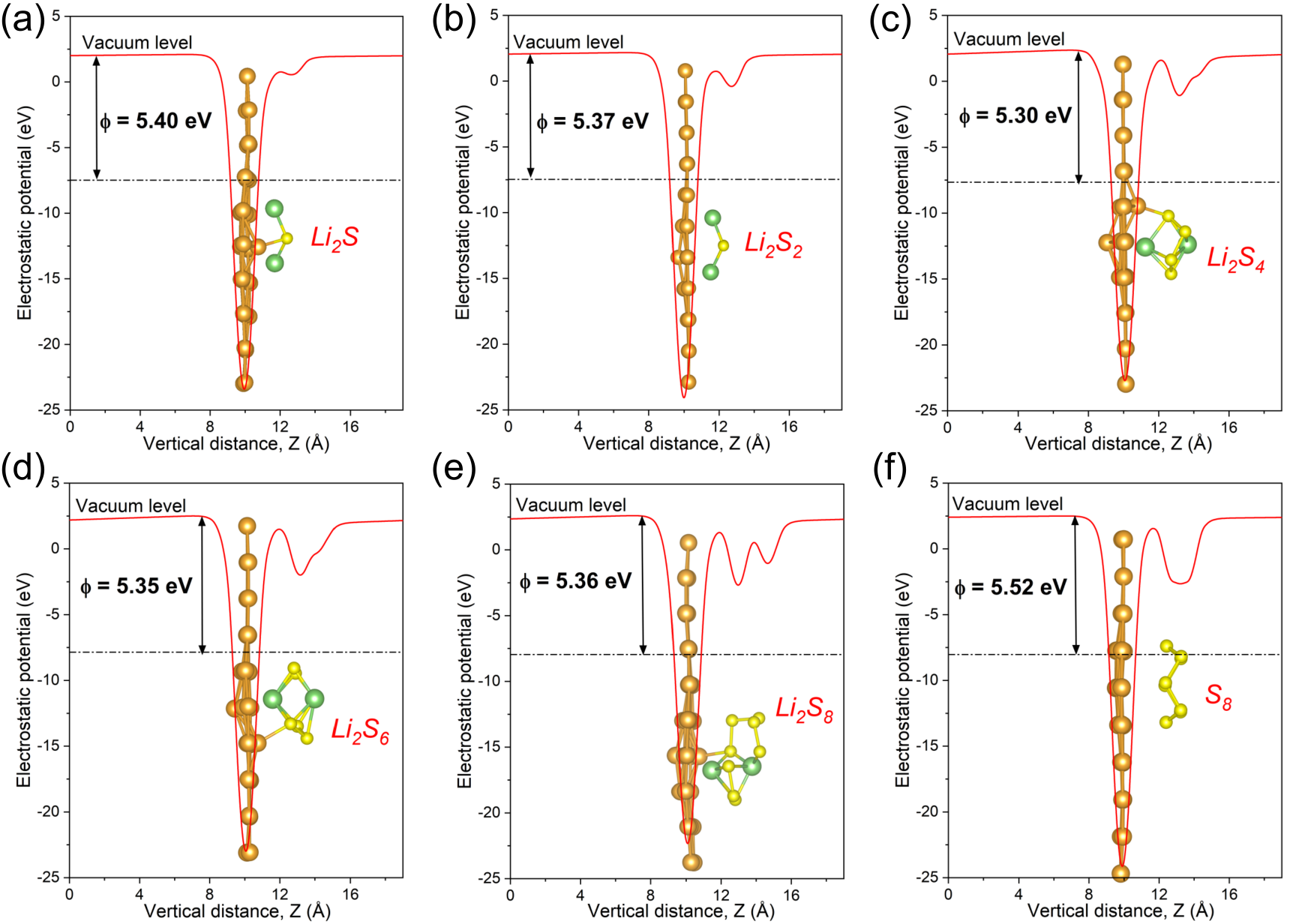}
    \caption{Planar-averaged electrostatic potential profiles along the surface normal for Goldene with adsorbed Li$_2$S$_n$ and S$_8$ clusters. Panels (a)–(f) correspond to Li$_2$S, Li$_2$S$_2$, Li$_2$S$_4$, Li$_2$S$_6$, Li$_2$S$_8$, and S$_8$, respectively. The work function ($\phi$) for pristine Goldene is found to be 5.50 eV.}
    \label{fig:7}
\end{figure*}

Another important descriptor for evaluating the modulation of a surface's reactivity is the work function (\(\Phi\)) of the substrate, which quantifies the minimum energy required to extract an electron from the Fermi level to the vacuum level \cite{martins2025high, elaggoune2025enhancing}. From a theoretical perspective, the work function is calculated according to
\begin{equation}
\Phi = V_{\infty} - E_F ,
\end{equation}
\noindent where $V_{\infty}$ represents the vacuum-level electrostatic potential and $E_F$ corresponds to the system's Fermi energy. Here, we assessed (\(\Phi\)) by computing the planar-averaged electrostatic potentials, where the distances were considered along the vertical \(z\)-direction with \(V(z) = \pm \infty\), as visualized in Figure \ref{fig:7}. Pristine Goldene exhibits a work function of 5.50~eV, reflecting its intrinsic metallic character. Upon introducing lithium polysulfides at the Goldene interface, pronounced asymmetric peaks emerge along the positive $z$-axis, indicating that the interaction with Li$_2$S$_n$ clusters perturbs the Goldene surface. Consequently, $\Phi$ decreases for all lithium polysulfide species, with reductions of 0.10, 0.13, 0.20, 0.15, and 0.14~eV for Li$_2$S, Li$_2$S$_2$, Li$_2$S$_4$, Li$_2$S$_6$, and Li$_2$S$_8$, respectively. This behavior illustrates the charge redistribution mechanisms discussed above. Lithium-based clusters donate electrons to the gold surface, thereby creating an interfacial layer that alters the electrostatic potential and reduces the work function. In contrast, due to the weak adsorption of S$_8$ on the Goldene monolayer, its work function is only slightly shifted to 5.52~eV.

\subsection{Catalytic potential of Goldene during charge-discharge reaction}

In this work, the sulfur reduction reaction (SRR) is examined by calculating the Gibbs free energy changes ($\Delta G$) along the charge–discharge pathway, which comprises a sequence of reactions from S$_8$ to the formation of Li$_2$S. This analysis is essential for improving the understanding of SRR kinetics, as a favorable free energy profile indicates thermodynamically viable Li–S bond cleavage and enhanced Li diffusion at the electrode interface.

\begin{figure*}
    \centering
    \includegraphics[width=0.85\linewidth]{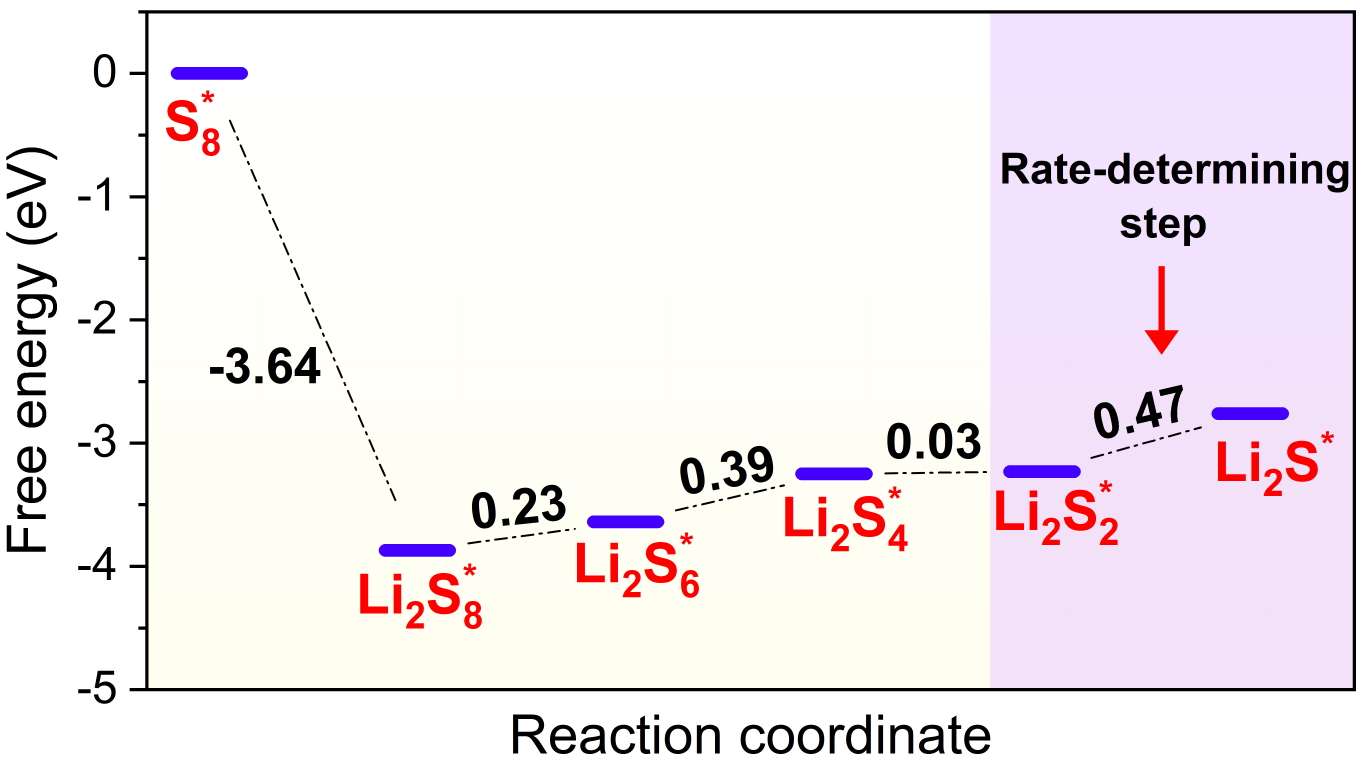}
    \caption{Reaction free energy profile for the stepwise lithiation of sulfur on the Goldene surface. The pathway illustrates the conversion from S$_8$ to Li$_2$S through successive polysulfide intermediates, Li$_2$S$_8$, Li$_2$S$_6$, Li$_2$S$_4$, and Li$_2$S$_2$. Energy values are referenced to the initial state, and the final Li$_2$S$_2$ to Li$_2$S transformation is identified as the rate determining step with a barrier of 0.47~eV.}
    \label{fig:8}
\end{figure*}

The catalytic kinetics of the sulfur reduction reaction (SRR) on Goldene are illustrated in Fig.~\ref{fig:8}. Negative $\Delta G$ values indicate spontaneous reaction steps, whereas positive $\Delta G$ values correspond to endothermic processes that require an external potential to proceed. The conversion from S$_8^{*}$ to Li$_2$S$_8^{*}$ is highly exothermic ($\Delta G = -3.64$~eV), which is expected due to electron injection in the initial reduction step, promoting the combination of Li atoms with sulfur chains. In contrast, the intermediate conversion steps are entirely endothermic, with energy barriers of 0.23~eV (Li$_2$S$_8^{*} \rightarrow$ Li$_2$S$_6^{*}$), 0.39~eV (Li$_2$S$_6^{*} \rightarrow$ Li$_2$S$_4^{*}$), and 0.03~eV (Li$_2$S$_4^{*} \rightarrow$ Li$_2$S$_2^{*}$). This trend is commonly observed and arises from the progressive cleavage of S–S bonds as the polysulfide chains shorten, a process that requires increasing energy input \cite{liu2022phosphorus, li2018revisiting}. Notably, the final conversion step to form Li$_2$S (Li$_2$S$_2^{*} \rightarrow$ Li$_2$S$^{*}$) exhibits the highest free-energy barrier, at $\Delta G = 0.47$~eV, identifying this step as the rate-determining step (RDS) of the SRR on Goldene.The maximum energy barrier observed on Goldene is lower than those reported for Mo$_2$CS$_2$ (0.83~eV) \cite{li2025adsorption}, Janus-type Ga$_2$SSe (0.94~eV), and Al$_2$SSe (1.05~eV) \cite{li2025janus}, as well as HDG-graphene (0.68~eV) \cite{wang2024unravelling}. These results confirm a moderate binding strength between lithium polysulfides and the Goldene monolayer, which enables effective anchoring while avoiding excessively strong adsorption that could lead to reaction hysteresis. In general, prohibitively high energy barriers during the SRR are indicative of strong chemisorption, which hampers polysulfide conversion by impeding desorption and slowing the overall charge--discharge process.

\section{Conclusion}

In summary, we systematically investigated the interactions between lithium polysulfide species and Goldene using first-principles calculations. All Li$_2$S$_n$ clusters exhibit strong adsorption on the Goldene monolayer, with adsorption energies ranging from $-4.29$ to $-1.90$~eV. These values are significantly stronger than the binding energies with common electrolyte solvents, demonstrating that Goldene is an effective substrate for anchoring lithium polysulfides. Charge-transfer analysis reveals substantial electron transfer from the Li-containing species to Goldene, reaching up to $0.92\,e$. This charge redistribution induces a strong interfacial dipole and reduces the work function of Goldene, indicating efficient modulation of surface reactivity upon lithium polysulfide adsorption.

Projected density of states (PDOS) calculations show that Goldene preserves its metallic character after adsorption. The Au $d$ states remain located at the Fermi level, and the pronounced hybridization between Au $d$ and S $p$ orbitals suggests moderate chemical bonding at the interface. Reaction free energy analysis further indicates that sulfur lithiation on Goldene is thermodynamically favorable. The overall stabilization energy reaches $-3.64$~eV, and the rate-determining barrier for the Li$_2$S$_2$ to Li$_2$S conversion is 0.47~eV, which is lower than that reported for conventional catalysts. This favorable Gibbs free energy profile suggests enhanced reaction kinetics for the sulfur reduction reaction (SRR) on the Goldene monolayer.

Overall, these results demonstrate that Goldene is a chemically active and electronically stable two-dimensional (2D) substrate capable of strongly binding lithium polysulfides and facilitating their structural transformation. Consequently, Goldene shows considerable promise for next-generation Li-S battery applications.

\section*{Data Availability}
All data supporting the findings of this study are available within the article.

\section*{Conflicts of interest}
\noindent The authors declare that they have no conflict of interest.

\section*{Acknowledgments}
This work was supported by the Brazilian funding agencies Fundação de Amparo à Pesquisa do Estado de São Paulo--FAPESP, grants 2024/21870-8, 2022/16509-9, and 2024/05087-1), and National Council for Scientific, Technological Development--CNPq, grant no. 307213/2021–8. L.A.R.J acknowledges financial support from FAPDF-PRONEM grant 00193.00001247 /2021-20, PDPG-FAPDF-CAPES Centro-Oeste grant number 00193-00000867/2024-94, and CNPq grants 301577/2025-0 and 444111/2024-7. 

\bibliographystyle{unsrtnat}
\bibliography{references}  

\end{document}